\documentclass[sigconf]{acmart}

\usepackage{algorithm}

\usepackage{graphicx} 
\usepackage{booktabs}
\usepackage{xcolor}
\usepackage{multirow}
\usepackage{makecell}
\usepackage{bm}
\usepackage{algpseudocode}  
\usepackage{amsmath}  
\usepackage{balance}
\usepackage[switch]{lineno}

\definecolor{cvprblue}{rgb}{0.21,0.49,0.74}
\definecolor{cYellow}{HTML}{FFFFCC}
\definecolor{cRed}{HTML}{FFCCCC} 
\definecolor{cGrey}{HTML}{F3F7F2} 
\definecolor{cGreen}{HTML}{339933}

\usepackage[dvipsnames]{xcolor}
\usepackage{colortbl}
\usepackage{subfig}
\usepackage{pifont}

\AtBeginDocument{%
  }

\copyrightyear{2025}
\acmYear{2025}
\setcopyright{acmlicensed}\acmConference[SIGIR '25]{Proceedings of the 48th International ACM SIGIR Conference on Research and Development in Information Retrieval}{July 13--18, 2025}{Padua, Italy}
\acmBooktitle{Proceedings of the 48th International ACM SIGIR Conference on Research and Development in Information Retrieval (SIGIR '25), July 13--18, 2025, Padua, Italy}
\acmDOI{10.1145/3726302.3730066}
\acmISBN{979-8-4007-1592-1/2025/07}

\begin{document}

\title{Queries Are Not Alone: Clustering Text Embeddings for Video Search}

\author{Peiyang Liu}
\affiliation{%
  \institution{National Engineering Research Center for Software Engineering, Peking University}
  \city{Beijing}
  \country{China}}
\email{liupeiyang@pku.edu.cn}

\author{Xi Wang}
\affiliation{%
  \institution{Peking University}
  \city{Beijing}
  \country{China}}
\email{wangxi5629@pku.edu.cn}

\author{Ziqiang Cui}
\affiliation{%
  \institution{City University of Hong Kong}
  \city{Hong Kong SAR}
  \country{China}}
\email{ziqiang.cui@my.cityu.edu.hk}

\author{Wei Ye}
\authornote{Corresponding Author}
\affiliation{%
  \institution{National Engineering Research Center for Software Engineering, Peking University}
  \city{Beijing}
  \country{China}}
\email{wye@pku.edu.cn}

\renewcommand{\shortauthors}{Peiyang Liu, Xi Wang, Ziqiang Cui, and Wei Ye}


\begin{abstract}
The rapid proliferation of video content across various platforms has highlighted the urgent need for advanced video retrieval systems. Traditional methods, which primarily depend on directly matching textual queries with video metadata, often fail to bridge the semantic gap between text descriptions and the multifaceted nature of video content. This paper introduces a novel framework, the Video-Text Cluster (VTC), which enhances video retrieval by clustering text queries to capture a broader semantic scope. 
We propose a unique clustering mechanism that groups related queries, enabling our system to consider multiple interpretations and nuances of each query. This clustering is further refined by our innovative Sweeper module, which identifies and mitigates noise within these clusters. Additionally, we introduce the Video-Text Cluster-Attention (VTC-Att) mechanism, which dynamically adjusts focus within the clusters based on the video content, ensuring that the retrieval process emphasizes the most relevant textual features.
Further experiments have demonstrated that our proposed model surpasses existing state-of-the-art models on five public datasets.
\end{abstract}




\maketitle

\section{Introduction}
The surge in video content across platforms has made video search essential for applications like multimedia search engines, digital libraries, and content recommendation systems \cite{cui2025semantic}. However, retrieving relevant videos from textual descriptions is challenging due to the semantic gap between text queries and the complex information in videos \cite{li2019w2vv++,gorti2022x,wu2023cap4video,li2023progressive,yang2024dgl,DBLP:conf/wsdm/LiSZXCCTWY24}.

Traditional systems often rely on direct matching of text queries with video metadata or annotations \cite{smoliar1994content,chang1997videoq,smith1997image,sivic2003video,jiang2007towards}, which fails to capture the deeper semantics of videos, as they handle only surface-level or explicitly tagged information. The brevity and ambiguity of text queries further complicate retrieval, making it hard to encompass the full video content.
Recent machine learning \cite{liu2024unsupervised} advancements, especially in deep learning \cite{liu2020not}, offer new ways to bridge the text-video semantic gap. Embedding-based retrieval systems \cite{liu2021distilling,liu2021improving,liu2022label} show promise by mapping text and video data into a shared semantic space for effective similarity assessment \cite{fang2023uatvr}.
These methods leverage neural networks to learn rich representations of text and video, enabling a more nuanced understanding and comparison of their content \cite{DBLP:conf/cvpr/WangZ021,zhao2022centerclip,huang2023vop,zala2023hierarchical,ventura2024covr,Wang_2024_CVPR}.

\begin{figure}
	    \centering
	    \includegraphics[scale=0.34]{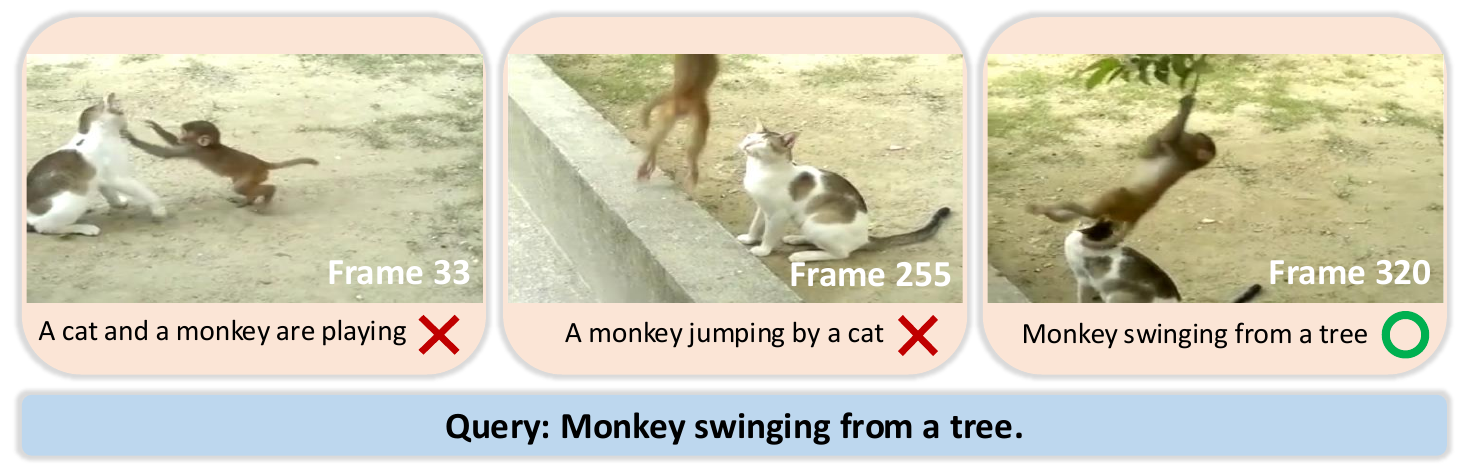}
     \vspace{-7mm}
	    \caption{A text-video pair in the MSRVTT dataset. A brief, solitary query struggles to adequately capture the complex semantics of the video.}
	    \label{fig:cat_case}
     \vspace{-5mm}
\end{figure} 

Despite these advancements, video retrieval continues to pose significant challenges. As illustrated in Figure \ref{fig:cat_case}, the dynamic nature of videos, characterized by sequences of diverse scenes and actions, introduces complexities absent in static images or straightforward text descriptions. A single video might contain multiple scenes, transitions, and temporal developments that are difficult to capture in a concise text query. For example, a video titled ``Monkey and cat'' might include scenes of the ``A cat and a monkey are playing'', ``A monkey jumping by a cat'', and ``Monkey swinging from a tree'' – a diversity of content that a simple query cannot fully express.
Furthermore, the contextual and often subjective interpretation of video content can differ greatly among viewers, complicating the development of a universally effective search system. What one user considers a ``thrilling action sequence'' might be perceived as ``excessive violence'' by another, highlighting the need for retrieval systems that can understand and accommodate these varying perspectives.
Current embedding-based methods also struggle with the ``one-to-many'' relationship between queries and relevant videos. A single query like ``dog playing'' could match videos of different dog breeds, various play activities (fetching, swimming, running), and diverse environments (park, beach, home), each representing valid but visually distinct matches. Existing approaches often fail to capture this semantic breadth, instead focusing narrowly on the literal interpretation of the query.

To address these challenges, we propose a novel approach called Video-Text Cluster, which enhances the semantic understanding of textual queries by considering them within their broader context. Rather than relying solely on individual text queries, our method clusters related queries to capture a wider range of semantic meanings. This clustering enables the system to consider multiple interpretations and nuances of the query, thereby providing a richer set of features for matching with video content.

For example, given a query ``cat playing'', our system would cluster it with semantically related phrases such as ``kitten with toy'', ``feline jumping'', and ``pet having fun''. This expanded semantic field allows our system to match videos that might not contain the exact words ``cat'' and ``playing'' but still represent the core concept the user is searching for. By expanding the semantic scope of the query through clustering, we create a more robust representation that can better match the visual complexity and diversity of video content.

Methods that enhance clustering inevitably introduce some noise. For instance, the cluster for ``cat playing'' might incorrectly include phrases like ``dog chasing ball'' or ``children at playground'' that share some semantic elements but represent different visual concepts. Our model incorporates a module called Sweeper, which identifies noise by analyzing the semantic relationship between the text within the cluster and the query. The Sweeper evaluates each element in the cluster and assigns a relevance score, allowing the system to prioritize the most semantically aligned phrases.

However, relying solely on semantics to eliminate noise may inadvertently remove valuable signals. Texts describing different frames of the same target video can possess entirely different semantics. For example, a video of ``a family vacation'' might contain scenes described as ``children building sandcastles'', ``parents relaxing under umbrellas'', and ``sunset over the ocean'' – all valid descriptions of the same video despite their semantic differences. To address this, we propose a novel Video-Text Cluster-Attention mechanism (VTC-Att). This mechanism integrates signals from video frames with the text semantic signals provided by the Sweeper to effectively clean the noise. By examining how well each clustered text matches with different frames of the video, VTC-Att can retain semantically diverse but visually relevant descriptions. This attention-based approach ensures that the system focuses on the most salient features, thereby improving the accuracy and relevance of the retrieved videos. The main contributions of our work are as follows:

\begin{itemize}
    \item We reevaluate the video retrieval framework and propose an innovative retrieval scheme centered on text clusters. Our approach provides more extensive semantic information for text embeddings, thereby bridging the semantic gap between the query and the video.
    \item We have designed a Sweeper module, which can effectively identify noise in text clusters based on semantic information.
    Furthermore, we have developed an innovative attention mechanism called VTC-Att. This mechanism effectively utilizes both the semantic signals provided by the Sweeper and the frame signals in the video to identify noise within text clusters.
    \item We conduct a comprehensive evaluation of the proposed method using five widely-used datasets: MSRVTT, LSMDC, DiDeMo, VATEX, and Charades. The experimental results indicate that our method consistently outperforms state-of-the-art retrieval models across all datasets.
\end{itemize}

\section{Related Work}

The field of video retrieval has witnessed remarkable advancements, largely propelled by pre-trained multimodal Transformers, particularly the CLIP model \cite{vaswani2017attention, radford2021learning}. While CLIP was initially designed for image-text retrieval, its foundational principles have proven naturally extensible to video retrieval tasks. This section systematically examines related work through two interconnected perspectives: (1) innovations in attention mechanisms and token manipulation strategies, and (2) adaptations of CLIP specifically tailored for video retrieval challenges. These complementary research streams have evolved synergistically, collectively driving methodological progress in the field.

\subsection{Attention Mechanisms and Token Manipulation}

Recent research has introduced sophisticated approaches to enhance representation learning through novel attention mechanisms and token manipulation strategies. TS2-Net \cite{liu2022ts2} pioneered a dual-pronged approach combining token shift and selection architectures, significantly optimizing the representation of temporal video content. Building on this foundation, X-Pool \cite{gorti2022x} leveraged cross-modal attention mechanisms to facilitate more effective information integration across modalities, resulting in substantial improvements in retrieval accuracy. Further advancing this trajectory, UATVR \cite{fang2023uatvr} introduced a comprehensive framework for aggregating multi-grained semantics, thereby enhancing reasoning capabilities and overall retrieval performance.

\subsection{CLIP Adaptations for Video Retrieval}

Concurrent with advances in attention mechanisms, researchers have developed increasingly sophisticated adaptations of CLIP for video retrieval tasks. CLIPBERT \cite{lei2021less} introduced a computationally efficient end-to-end learning strategy utilizing sparse sampling techniques, establishing an important balance between performance and computational cost. CLIP4Clip \cite{luo2022clip4clip} further refined this approach with specific optimizations for video retrieval tasks, while CenterCLIP \cite{zhao2022centerclip} implemented an innovative multi-segment frame clustering algorithm to enhance precision. 

As the field progressed, more specialized adaptations emerged. CLIP-ViP \cite{xue2022clip} addressed critical challenges related to data scale and domain adaptation, while Cap4video \cite{wu2023cap4video} introduced groundbreaking zero-shot video captioning capabilities. TEFAL \cite{ibrahimi2023audio} expanded the multimodal landscape by aligning audio and video with text queries, enabling more comprehensive retrieval capabilities. More recent innovations include T-MASS \cite{Wang_2024_CVPR}, which employs stochastic text embeddings to enhance accuracy, DiffusionRet \cite{jin2023diffusionret}, which integrates generative and discriminative methodologies, and DGL \cite{yang2024dgl}, which refines retrieval through cross-modal dynamic prompt tuning and Global-Local video attention mechanisms.

Our work builds upon these significant advancements by leveraging a CLIP-based multimodal transformer architecture while fundamentally revisiting the video retrieval framework. We introduce a novel methodological approach centered on text cluster embeddings, offering a distinctive contribution to the field's ongoing evolution. This approach not only synthesizes insights from prior research but also charts new directions for enhancing video retrieval performance in alignment with state-of-the-art standards.

\begin{figure*}
	    \centering
	    \includegraphics[scale=0.5]{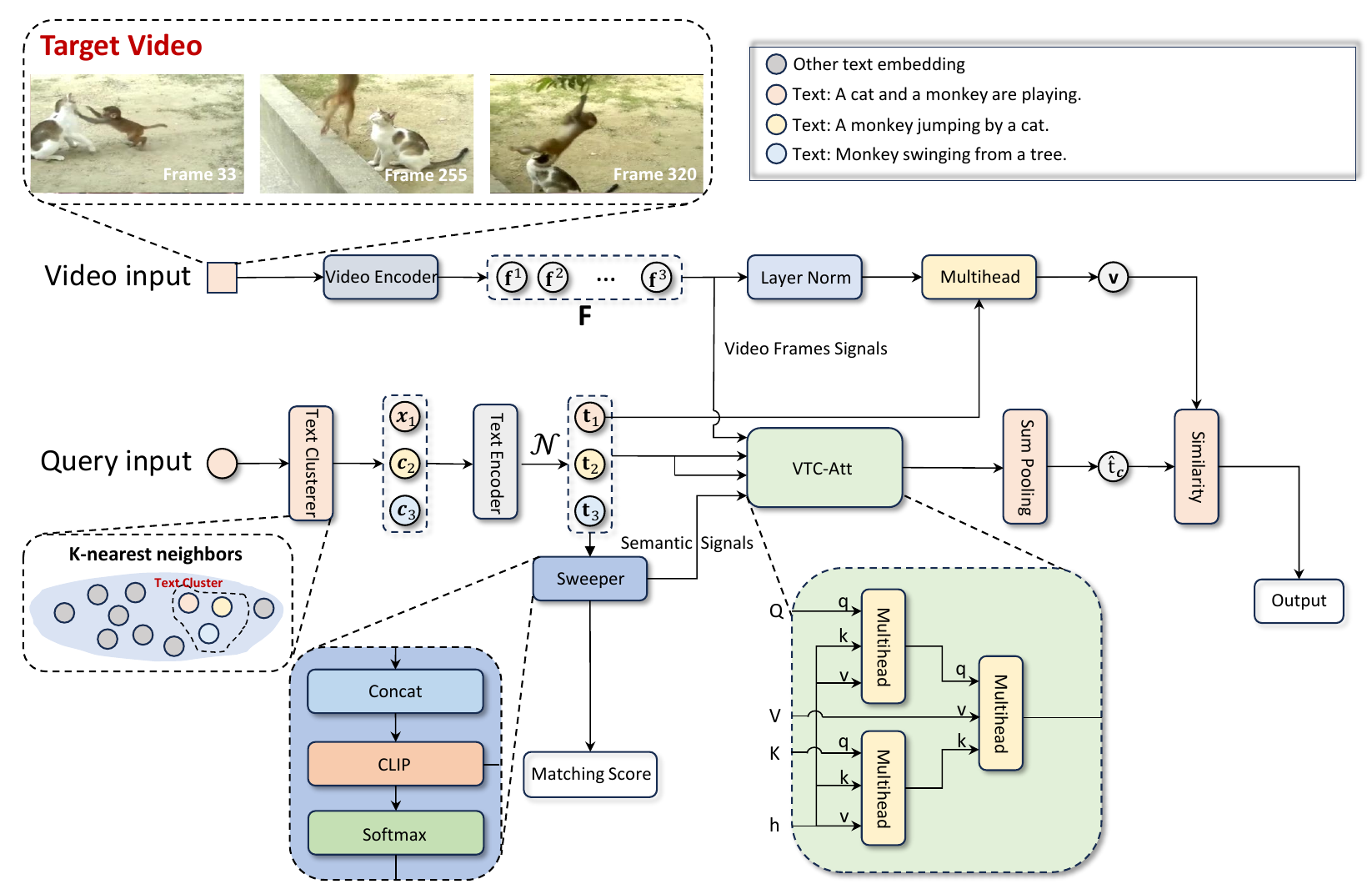}
     \vspace{-3mm}
        \caption{
        Overview of our proposed method. 
        The Text Clusterer groups all texts into distinct clusters. Subsequently, the Text Encoder and Video Encoder encode these clustered texts and video frames. The Sweeper identifies noise within the clustered texts and generates semantic signals. The VTC-Att mechanism then combines these semantic signals with video frame signals to eliminate noise. Ultimately, the refined text cluster embeddings are employed for retrieval.
        }
	    \label{fig:method}
     \vspace{-5mm}
\end{figure*} 

\section{Methodology}

\subsection{Preliminary}
Let $\Omega = (\mathcal{D}, \mathcal{V}, f)$ be a triple where $\mathcal{D}=\{x_i\}_{i=1}^N$ denotes the text corpus consisting of $N$ discrete textual elements, and $\mathcal{V}=\{y_i\}_{i=1}^M$ represents the video collection containing $M$ video sequences. The mapping $f: \mathcal{D} \times \mathcal{V} \rightarrow \mathbb{R}^+$ defines the relevance scoring function between text-video pairs.
The fundamental objective of our video retrieval framework can be formalized as finding an optimal mapping function $f^*$ such that:
\begin{equation}
f^* = \mathrm{argmax}_{f \in \hat{\mathcal{F}}} \sum_{x_i \in \mathcal{D}} \sum_{y_j \in \mathcal{V}} \Phi(x_i, y_j) \cdot \mathbb{I}[f(x_i) = y_j],
\end{equation}
where $\hat{\mathcal{F}}$ represents the hypothesis space of all possible mapping functions, and $\mathbb{I}[\cdot]$ is the indicator function.

To solve this optimization problem, we first project both modalities into a shared embedding space $\mathbb{R}^o$ through the embedding functions $\Phi_\zeta: \mathcal{D} \rightarrow \mathbb{R}^o$ and $\Phi_\xi: \mathcal{V} \rightarrow \mathbb{R}^o$, resulting in text embeddings $\mathbf{t}_i = \Phi_\zeta(x_i)$ and video embeddings $\mathbf{v}_i = \Phi_\xi(y_i)$, where $o$ denotes the dimensionality of the embedding space.

The overall framework of our proposed approach is depicted in Figure \ref{fig:method}.
Our approach mainly includes Six structures: Text Clusterer, Text Encoder, Video Encoder, Sweeper, VTC-Att, and Retriever. Firstly, we use the Text Clusterer to cluster all texts, thereby obtaining text clusters. Next, we use the Text Encoder to encode the query and texts in the text cluster, while using the Video Encoder to encode the video. Then, we use the Sweeper to perform noise recognition on the text cluster and utilize the VTC-Att mechanism to clean the noise.
Finally, Retriever use the text cluster embeddings and video embeddings for video retrieval.

\subsection{Text Clusterer}

Let $\Phi_{C}: \mathcal{D} \rightarrow \mathbb{R}^o$ denote the CLIP text encoder function of Text Clusterer operating on input space $\mathcal{D}$. For any given text input $x_i \in \mathcal{D}$, the embedding process can be formalized as:
\begin{equation}
\mathbf{t}^*_i = \Phi_{C}(x_i) \in \mathbb{R}^o.
\end{equation}
To optimize the semantic clustering objective, we introduce a novel contrastive learning framework \cite{liu2021quadrupletbert,liu2023retrieval}. For each text $x_i$, we construct a triplet:
\begin{equation}
\tau_i = (x_i, p_i, q_i) \in \mathcal{T} \subset \mathcal{D} \times \mathcal{D} \times \mathcal{D},
\end{equation}
where $\mathcal{T}$ represents the triplet space. The construction of positive and negative samples follows a stochastic dropout-based approach. Let $\mathcal{Z}$ be the space of all possible dropout \cite{baldi2013understanding} masks, and $z, z' \sim \mathcal{Z}$ be independently sampled masks. The embedding process with dropout can be formalized as:
\begin{equation}
\begin{aligned}
\Phi_{C}(x, z): \mathcal{D} \times \mathcal{Z} &\rightarrow \mathbb{R}^o \\
(x, z) &\mapsto \mathbf{t}^*_{x,z}.
\end{aligned}
\end{equation}
The triplet construction process can then be expressed as:
\begin{equation}
\begin{cases}
\mathbf{t}^*_{x_i} = \Phi_{C}(x_i, z), \\
\mathbf{t}^*_{p_i} = \Phi_{C}(x_i, z'), \\
\mathbf{t}^*_{q_i} = \Phi_{C}(\varsigma(D\setminus\{x_i\})),
\end{cases}
\end{equation}
where $\varsigma$ is a random sampling function.
The loss function $\mathcal{L}_C$ incorporates a margin-based contrastive objective:
\begin{equation}
\mathcal{L}_C(\tau_i) = \max(0, \mathcal{U}(\mathbf{t}^*_{x_i}, \mathbf{t}^*_{p_i}) - \mathcal{U}(\mathbf{t}^*_{x_i}, \mathbf{t}^*_{q_i}) + \gamma),
\end{equation}
where $\gamma \in \mathbb{R}^+$ represents the margin constant, and $\mathcal{U}: \mathbb{R}^o \times \mathbb{R}^o \rightarrow \mathbb{R}$ is the cosine distance metric defined as:
\begin{equation}
\mathcal{U}(\mathbf{X}, \mathbf{Y}) = 1 - \frac{\langle \mathbf{X}, \mathbf{Y} \rangle}{\|\mathbf{X}\|_2 \cdot \|\mathbf{Y}\|_2},
\end{equation}
where $\langle \cdot, \cdot \rangle$ denotes the inner product and $\|\cdot\|_2$ represents the L2 norm.

\subsection{Text and Video Encoders}

Let $\Omega_{MM} = (\Phi_\xi, \Phi_\zeta)$ represent our multimodal encoding framework, where $\Phi_\xi$ and $\Phi_\zeta$ denote the CLIP image and text encoders respectively. For a video sequence $y_i = \{f_i^t\}_{t=1}^T \in \mathcal{V}$, where $T$ represents the temporal dimension, we define a temporal sampling function $\Theta: \mathbb{R}^{T} \rightarrow \mathbb{R}^{T'}$ that extracts $T'$ representative frames.
The multimodal encoding process can be formalized as:
\begin{equation}
\begin{aligned}
\mathbf{F}_i &= [\mathbf{f}_i^1, \mathbf{f}_i^2, ..., \mathbf{f}_i^{T'}] \in \mathbb{R}^{T' \times o}, \\
\mathbf{f}_i^j &= \Phi_\xi(f_i^j)\\
\mathbf{t}_i &= \Phi_\zeta(x_i) \in \mathbb{R}^o,
\end{aligned}
\end{equation}
where $\mathbf{F}_i$ represents the stacked frame embeddings and $\mathbf{t}_i$ denotes the text embedding.

\subsection{Sweeper}
\label{sec: sweeper}
For a given text $x_i$ and its corresponding cluster $\mathcal{C}_i$ obtained through Approximate Nearest Neighbors (ANN) \cite{DBLP:conf/nips/Shrivastava014,DBLP:conf/aistats/GuoKCS16}, we construct an augmented sequence through a hierarchical composition function:

\begin{equation}
\begin{aligned}
\hat{x}_i &= \omega(\xi_{cls} \oplus x_i \oplus \xi_{sep} \oplus \Gamma(\mathcal{C}_i)), \\
\mathcal{C}_i &= \{c_j | c_j \in \Lambda_\mathrm{ANN}(x_i, \mathcal{K}), j \neq i\},
\end{aligned}
\end{equation}
where:
\begin{itemize}
    \item $\omega$ is a sequence concatenation operator.
    \item $\xi_{cls}, \xi_{sep}$ are special tokens [CLS] and [SEP].
    \item $\Gamma$ is a cluster aggregation function.
    \item $\oplus$ denotes element-wise concatenation.
    \item $\Lambda_\mathrm{ANN}(\cdot, \mathcal{K})$ represents the approximate $\mathcal{K}$-nearest neighbors function, which can obtain the approximate nearest $\mathcal{K}$ neighbors from the embedding space at high speed within milliseconds.
\end{itemize}

\subsubsection{Cross-attention Feature Extraction}

The feature extraction process employs a specialized CLIP text encoder $\Phi_s$ of Sweeper with cross-attention mechanisms:
\begin{equation}
\mathbf{h}_i = \Phi_s(\hat{x}_i) \in \mathbb{R}^d,
\end{equation}
where the $\Phi_s$ consists of $L$ layers of self-attention and feed-forward networks, $d$ is the hidden size.

\subsubsection{Noise Detection Classification}

The classification implements a probabilistic mapping from feature space to segment probabilities $\mathbf{s}_i$:
\begin{equation}
\mathbf{s}_i = \sigma(\mathbf{W}_s\mathbf{h}_i + \mathbf{b}_s),
\end{equation}
where:
\begin{itemize}
    \item $\mathbf{W}_s \in \mathbb{R}^{g \times d}$ is the weight matrix
    \item $\mathbf{b}_s \in \mathbb{R}^g$ is the bias vector
    \item $\sigma$ is the softmax function
    \item $g$ is the number of noise segments
\end{itemize}

\subsubsection{Automatic Label Generation}

We introduce an automatic label generation mechanism based on Jaccard similarity \cite{ferdous2009efficient} segmentation. Let $\beta: \mathcal{D} \times \mathcal{D} \rightarrow [0,1]$ be the Jaccard similarity function. For each text pair $(x_i, c_j)$ where $c_j \in \mathcal{C}_i$, we compute:

\begin{equation}
\begin{aligned}
B_{ij} &= \beta(x_i, c_j), \\
\mathcal{S}_k &= \{(x_i, c_j) | \frac{k-1}{g} \leq B_{ij} < \frac{k}{g}\}, k \in \{1,\ldots,g\}.
\end{aligned}
\end{equation}
The label vector $\mathbf{y}_i$ for text $c_j$ is then constructed as:
\begin{equation}
\mathbf{y}_j[k] = \begin{cases}
1 & \text{if } (x_i, c_j) \in \mathcal{S}_k \text{ for some } c_j \in \mathcal{C}_i, \\
0 & \text{otherwise}.
\end{cases}
\end{equation}

\subsubsection{Loss Function}
The Sweeper is trained using a weighted cross-entropy loss with label smoothing:
\begin{equation}
\mathcal{L}_{\mathrm{sweep}} = -\sum_{i=1}^N \sum_{k=1}^g \omega_k((1-\epsilon)\mathbf{y}_i[k] + \frac{\epsilon}{g})\log(\mathbf{s}_i[k]),
\end{equation}
where:
\begin{itemize}
    \item $\omega_k$ is the class-specific weight to handle class imbalance:
    \begin{equation}
    \omega_k = \frac{N}{\sum_{i=1}^N \mathbf{y}_i[k]} \cdot \frac{1}{g}
    \end{equation}
    \item $\epsilon$ is the label smoothing factor
    \item $N$ is the total number of training samples
\end{itemize}

\subsection{Video-Text-Cluster Attention}
For each text $x_i$, we construct a neighborhood matrix $\mathcal{N}_i \in \mathbb{R}^{(\mathcal{K}+1) \times o}$ through the following mapping:
\begin{equation}
\mathcal{N}_i = \begin{bmatrix} 
\mathbf{t}_i \\
\Phi_\zeta(c_1) \\
\vdots \\
\Phi_\zeta(c_\mathcal{K})
\end{bmatrix}, \quad c_j \in \Lambda_{ANN}(x_i, \mathcal{K}).
\end{equation}

It is evident that while cluster-based methods introduce richer semantics, they may also introduce additional noise, which, if not addressed, could negatively impact model performance. As described in Section \ref{sec: sweeper}, the Sweeper can provide semantic signals for noise reduction. However, relying solely on the semantic relevance between the query and texts is insufficient. Due to the complexity of video semantics, some texts that are not particularly semantically related may actually describe different frames of the same video.
To eliminate the noise signals introduced into the text cluster, our proposed method utilizes the signals from each video frame combined with the additional semantic signals provided by the Sweeper for noise identification. Specifically, we aim to assign a higher weight to a text within the text cluster that adequately describes the video and a lower weight when it is not significantly related to the video. Based on this motivation, we propose the Video-Text Cluster Attention (VTC-Att) mechanism, which is an extension of the multihead attention mechanism \cite{vaswani2017attention}.
As shown in Figure \ref{fig:method}, our VTC-Att includes four input components: Q (query), K (key), V (value), and h (semantic signals from Sweeper). We employ two additional multihead attention modules to integrate the semantic signals from Sweeper into the query and key:
\begin{equation}
\begin{aligned}
\mathbf{Q} &= \text{LayerNorm}(\text{MultiHead}(\mathbf{F}_i\mathbf{W}_Q, \mathbf{h}_i\mathbf{W}_K, \mathbf{h}_i\mathbf{W}_V)), \\
\mathbf{K} &= \text{LayerNorm}(\text{MultiHead}(\mathcal{N}_i\mathbf{W}_Q', \mathbf{h}_i\mathbf{W}_K', \mathbf{h}_i\mathbf{W}_V')),
\end{aligned}
\end{equation}
where $\mathbf{W}_{\{\cdot\}} \in \mathbb{R}^{d \times d_k}$ are learnable projection matrices, and $\text{LayerNorm}(\cdot)$ is the Layernorm function \cite{xu2019understanding}.
The final attention output is computed as:
\begin{equation}
\hat{\mathbf{t}}_i = \text{LayerNorm}(\text{MultiHead}(\mathbf{Q}\mathbf{W}_Q'', \mathbf{K}\mathbf{W}_K'', \mathcal{N}_i\mathbf{W}_V'')).
\end{equation}

\subsection{Text-Video Feature Fusion}
To create comprehensive video embeddings from the embeddings of individual frames, we utilize the Transformer model \cite{vaswani2017attention} as our feature extractor. This choice is motivated by the fact that not every frame in a video is pertinent to the accompanying text. The Transformer employs an attention mechanism that assigns greater weights to significant frames while reducing the weights of irrelevant ones. This mechanism allows the model to concentrate more effectively on frames that are relevant to the text. The fusion process can be mathematically represented as a series of attention operations:
\begin{equation}
\begin{aligned}
\mathbf{v}_i &= \Psi_{\mathrm{fusion}}(\mathbf{t}_i, \mathbf{F}_i) \\
&= \text{LayerNorm}(\text{MultiHead}(\mathbf{t}_i\mathbf{W}^*_Q, \mathbf{F}_i\mathbf{W}^*_K, \mathbf{F}_i\mathbf{W}^*_V)).
\end{aligned}
\end{equation}

\begin{algorithm}
\caption{Video Retrieval with Text Clustering and Noise Reduction}
\begin{algorithmic}[1]
\Require Text corpus $\mathcal{D}=\{x_i\}_{i=1}^N$, Video collection $\mathcal{V}=\{y_i\}_{i=1}^M$
\Ensure Optimized mapping function $f^*$ for text-video retrieval

\State \textbf{/* Text Clusterer */}
\For{each text $x_i \in \mathcal{D}$}
    \State $\mathbf{t}^*_i \gets \Phi_{C}(x_i)$ \Comment{Encode text using CLIP text encoder}
    \State Sample dropout masks $z, z' \sim \mathcal{Z}$
    \State $\mathbf{t}^*_{x_i} \gets \Phi_{C}(x_i, z)$
    \State $\mathbf{t}^*_{p_i} \gets \Phi_{C}(x_i, z')$ \Comment{Positive sample with different dropout}
    \State $\mathbf{t}^*_{q_i} \gets \Phi_{C}(\varsigma(D\setminus\{x_i\}))$ \Comment{Random negative sample}
    \State $\mathcal{L}_C(\tau_i) \gets \max(0, \mathcal{U}(\mathbf{t}^*_{x_i}, \mathbf{t}^*_{p_i}) - \mathcal{U}(\mathbf{t}^*_{x_i}, \mathbf{t}^*_{q_i}) + \gamma)$
\EndFor
\State Optimize Text Clusterer using $\mathcal{L}_C$

\State \textbf{/* Text and Video Encoders */}
\For{each text $x_i \in \mathcal{D}$}
    \State $\mathbf{t}_i \gets \Phi_\zeta(x_i)$ \Comment{Encode text using CLIP text encoder}
\EndFor
\For{each video $y_i \in \mathcal{V}$}
    \State $\{f_i^j\}_{j=1}^{T'} \gets \Theta(y_i)$ \Comment{Sample $T'$ frames from video}
    \For{$j = 1$ to $T'$}
        \State $\mathbf{f}_i^j \gets \Phi_\xi(f_i^j)$ \Comment{Encode each frame using CLIP image encoder}
    \EndFor
    \State $\mathbf{F}_i \gets [\mathbf{f}_i^1, \mathbf{f}_i^2, ..., \mathbf{f}_i^{T'}]$ \Comment{Stack frame embeddings}
\EndFor

\State \textbf{/* Sweeper */}
\For{each text $x_i \in \mathcal{D}$}
    \State $\mathcal{C}_i \gets \Lambda_\mathrm{ANN}(x_i, \mathcal{K})$ \Comment{Find $\mathcal{K}$ nearest neighbors}
    \State $\hat{x}_i \gets \omega(\xi_{cls} \oplus x_i \oplus \xi_{sep} \oplus \Gamma(\mathcal{C}_i))$ \Comment{Construct augmented sequence}
    \State $\mathbf{h}_i \gets \Phi_s(\hat{x}_i)$ \Comment{Extract features using specialized CLIP encoder}
    \State $\mathbf{s}_i \gets \sigma(\mathbf{W}_s\mathbf{h}_i + \mathbf{b}_s)$ \Comment{Noise detection classification}
    
    \For{each $c_j \in \mathcal{C}_i$}
        \State $B_{ij} \gets \beta(x_i, c_j)$ \Comment{Compute Jaccard similarity}
        \For{$k = 1$ to $g$}
            \If{$\frac{k-1}{g} \leq B_{ij} < \frac{k}{g}$}
                \State $\mathbf{y}_j[k] \gets 1$ \Comment{Generate automatic label}
            \Else
                \State $\mathbf{y}_j[k] \gets 0$
            \EndIf
        \EndFor
    \EndFor
    
    \For{$k = 1$ to $g$}
        \State $\omega_k \gets \frac{N}{\sum_{i=1}^N \mathbf{y}_i[k]} \cdot \frac{1}{g}$ \Comment{Compute class weights}
    \EndFor
    
    \State $\mathcal{L}_{\mathrm{sweep}} \gets -\sum_{i=1}^N \sum_{k=1}^g \omega_k((1-\epsilon)\mathbf{y}_i[k] + \frac{\epsilon}{g})\log(\mathbf{s}_i[k])$ \Comment{Compute loss}
\EndFor
\end{algorithmic}
\end{algorithm}

\begin{algorithm}
\ContinuedFloat
\caption{Video Retrieval with Text Clustering and Noise Reduction (continue)}
\label{algo:pesudo}
\begin{algorithmic}[1]
\State \textbf{/* Video-Text-Cluster Attention */}
\For{each text $x_i \in \mathcal{D}$}
    \State $\mathcal{N}_i \gets [\mathbf{t}_i; \Phi_\zeta(c_1); \ldots; \Phi_\zeta(c_\mathcal{K})]$ where $c_j \in \Lambda_{ANN}(x_i, \mathcal{K})$ \Comment{Construct neighborhood matrix}
    \State $\mathbf{Q} \gets \text{LayerNorm}(\text{MultiHead}(\mathbf{F}_i\mathbf{W}_Q, \mathbf{h}_i\mathbf{W}_K, \mathbf{h}_i\mathbf{W}_V))$ \Comment{Query with semantic signals}
    \State $\mathbf{K} \gets \text{LayerNorm}(\text{MultiHead}(\mathcal{N}_i\mathbf{W}_Q', \mathbf{h}_i\mathbf{W}_K', \mathbf{h}_i\mathbf{W}_V'))$ \Comment{Key with semantic signals}
    \State $\hat{\mathbf{t}}_i \gets \text{LayerNorm}(\text{MultiHead}(\mathbf{Q}\mathbf{W}_Q'', \mathbf{K}\mathbf{W}_K'', \mathcal{N}_i\mathbf{W}_V''))$ \Comment{Apply attention to get refined text embedding}
\EndFor

\State \textbf{/* Text-Video Feature Fusion */}
\For{each video $y_i \in \mathcal{V}$}
    \State $\mathbf{v}_i \gets \text{LayerNorm}(\text{MultiHead}(\mathbf{t}_i\mathbf{W}^*_Q, \mathbf{F}_i\mathbf{W}^*_K, \mathbf{F}_i\mathbf{W}^*_V))$ \Comment{Fuse text and video features}
\EndFor

\State \textbf{/* Joint Training */}
\For{each batch $\mathcal{B}$ of size $S$}
    \State $\Lambda_{CBS}(\mathcal{B}) \gets \{(x_i, y_i, \{y_j\}_{j \neq i})\}_{i=1}^S$ \Comment{Cross-batch negative sampling}
    \State $\mathcal{L}_{\mathrm{ret}} \gets -\frac{1}{S}\sum_{i=1}^S \log \frac{\exp(\lambda \cdot \mathcal{U}(\hat{\mathbf{t}}_i, \mathbf{v}_i))}{\sum_{j=1}^S \exp(\lambda \cdot \mathcal{U}(\hat{\mathbf{t}}_j, \mathbf{v}_j))}$ \Comment{InfoNCE loss}
    \State $\mathcal{L}_{\mathrm{total}} \gets \mathcal{L}_{\mathrm{sweep}} + \mathcal{L}_{\mathrm{ret}}$ \Comment{Total loss}
    \State Update model parameters using $\mathcal{L}_{\mathrm{total}}$
\EndFor

\State \textbf{/* Inference */}
\For{query text $x_q$}
    \State $\hat{\mathbf{t}}_q \gets$ Process $x_q$ through Text Encoder, Sweeper, and VTC-Att
    \State $y^* \gets \mathrm{argmax}_{\mathbf{v}_j \in \mathcal{V}} \frac{\langle \hat{\mathbf{t}}_q, \mathbf{v}_j \rangle}{\|\hat{\mathbf{t}}_q\|_2 \cdot \|\mathbf{v}_j\|_2}$ \Comment{Retrieve most relevant video}
    \State \Return $y^*$
\EndFor

\end{algorithmic}
\end{algorithm}

\subsection{Joint Training}
We continue to adopt contrastive learning for training our Retriever. This approach requires not only the positive text-video pairs from the training set but also a large number of negative instances. Directly sampling negative video instances randomly could consume considerable GPU memory. Therefore, we employ the Cross-Batch Negative Sampling method $\Lambda_{CBS}$ \cite{wang2021cross}, allowing all instances within each batch to serve as negative instances for other instances in the batch, eliminating the need to construct additional negative instances. This approach only requires the regular loading of positive text-video pairs for each batch and significantly conserves GPU memory. 
For a batch $\mathcal{B}$ of size $S$, the sampling process can be formalized as:
\begin{equation}
\Lambda_{CBS}(\mathcal{B}) = \{(x_i, y_i, \{y_j\}_{j \neq i})\}_{i=1}^S.
\end{equation}
The retrieval objective is optimized through a temperature-scaled InfoNCE \cite{oord2018representation} loss:
\begin{equation}
    \mathcal{L}_{\mathrm{ret}} = -\frac{1}{S}\sum_{i=1}^S \log \frac{\exp(\lambda \cdot \mathcal{U}(\hat{\mathbf{t}}_i, \mathbf{v}_i))}{\sum_{j=1}^S \exp(\lambda \cdot \mathcal{U}(\hat{\mathbf{t}}_j, \mathbf{v}_j))}
\end{equation}
where $\lambda \in \mathbb{R}^+$ is a learnable temperature parameter that modulates the sharpness of the probability distribution. The total loss $\mathcal{L}_{\mathrm{total}}$ of joint training is the combination of $\mathcal{L}_{\mathrm{sweep}}$ and $\mathcal{L}_{\mathrm{ret}}$:
\begin{equation}
    \mathcal{L}_{\mathrm{total}} = \mathcal{L}_{\mathrm{sweep}} + \mathcal{L}_{\mathrm{ret}}.
\end{equation}
The retrieval function $\Psi_{ret}$ during inference can be defined as:
\begin{equation}
\Psi_{ret}(x_i, \mathcal{V}) = \mathrm{argmax}_{\mathbf{v}_j \in \mathcal{V}} \frac{\langle \hat{\mathbf{t}}_i, \mathbf{v}_j \rangle}{\|\hat{\mathbf{t}}_i\|_2 \cdot \|\mathbf{v}_j\|_2}.
\end{equation}
The pseudo code of VTC are defined in the Algorithm \ref{algo:pesudo}.

\section{Experiments}
\begin{table*}[t]
\begin{center}
\scalebox{0.93}{
\setlength{\tabcolsep}{3.4mm}{
\begin{tabular}{c|ccccc|ccccc} 
    \toprule 
     \multirow{2}{*}{Method} & \multicolumn{5}{c|}{MSRVTT Retrieval} & \multicolumn{5}{c}{LSMDC Retrieval} \\
     \cline{2-11}
     & R@1 \small{$\uparrow$} & R@5 \small{$\uparrow$} & R@10 \small{$\uparrow$} & MdR \small{$\downarrow$} & MnR \small{$\downarrow$} & R@1 \small{$\uparrow$} & R@5 \small{$\uparrow$} & R@10 \small{$\uparrow$} & MdR  \small{$\downarrow$} & MnR \small{$\downarrow$} \\
    \hline
    {\color{cGreen}{\color{cGreen}\textit{\small{CLIP-ViT-B/32}}}} & & & & & & & & & & \\ 
    X-Pool~\cite{gorti2022x}  & 46.9 & 72.8 & 82.2 & 2.0 & 14.3 & 25.2 & 43.7 & 53.5 & 8.0 & 53.2  \\
    DiffusionRet~\cite{jin2023diffusionret}   & 49.0 & 75.2 & 82.7 & 2.0 & 12.1 & 24.4 & 43.1 & 54.3 & 8.0 & 40.7 \\
     UATVR~\cite{fang2023uatvr}  & 47.5 & 73.9 & 83.5 & 2.0 & 12.3  & -- & -- & -- & -- & -- \\
    TEFAL~\cite{ibrahimi2023audio} &49.4 & 75.9 & 83.9& 2.0& 12.0& 26.8 &46.1 &56.5 &7.0 &44.4 \\
    CLIP-ViP~\cite{xue2022clip} & 50.1 & 74.8 & 84.6 & 1.0 & -- & 25.6 & 45.3 & 54.4 & 8.0 & --\\
    \cellcolor{cGrey}VTC (\small{Ours})  &\cellcolor{cGrey}\textbf{53.1} &\cellcolor{cGrey}\textbf{78.5}
    &\cellcolor{cGrey}\textbf{88.5} &\cellcolor{cGrey}\textbf{1.0} &\cellcolor{cGrey}\textbf{9.6} &\cellcolor{cGrey}\textbf{30.1} &\cellcolor{cGrey}\textbf{49.9} &\cellcolor{cGrey}\textbf{60.3} &\cellcolor{cGrey}\textbf{6.0} &\cellcolor{cGrey}\textbf{39.9}  \\
    \hline 
    {\color{cGreen}\textit{\small{CLIP-ViT-B/16}}} & & & & & & & & & & \\
    X-Pool~\cite{gorti2022x} &48.2 &73.7 &82.6 &2.0 &12.7 &26.1 &46.8 &56.7 &7.0 &47.3 \\
    UATVR~\cite{fang2023uatvr} & 50.8 & 76.3 & 85.5 & 1.0 & 12.4  & -- &-- &-- & --&-- \\ 
    CLIP-ViP~\cite{xue2022clip} & 54.2 & 77.2 & 84.8 &1.0 & -- & 29.4 & 50.6 & 59.0 & 5.0 & -- \\
    \cellcolor{cGrey}VTC (\small{Ours})  &\cellcolor{cGrey}\textbf{54.8} &\cellcolor{cGrey}\textbf{79.2} &\cellcolor{cGrey}\textbf{88.7} &\cellcolor{cGrey} \textbf{1.0} &\cellcolor{cGrey}\textbf{9.7} &\cellcolor{cGrey}\textbf{30.9} &\cellcolor{cGrey}\textbf{53.4} &\cellcolor{cGrey}\textbf{64.2} &\cellcolor{cGrey}\textbf{5.0} &\cellcolor{cGrey}\textbf{39.0}  \\
    \bottomrule
\end{tabular}
}
}
\end{center}
\caption{Text-to-video comparisons on MSRVTT~\cite{xu2016msr} and LSMDC~\cite{rohrbach2015dataset}. Bold denotes the best performance. 
``--'': result is unavailable.
} 
\label{tab: benchmark-MSRVTT9K-LSMDC}
\vspace{-8mm}
\end{table*}

\begin{table*}[t]
\begin{center}
\scalebox{0.93}{
\setlength{\tabcolsep}{3.4mm}{
\begin{tabular}{c|ccccc|ccccc} 
    \toprule 
     \multirow{2}{*}{Method} & \multicolumn{5}{c|}{DiDeMo Retrieval} & \multicolumn{5}{c}{VATEX Retrieval} \\
     \cline{2-11}
     & R@1 \small{$\uparrow$} & R@5 \small{$\uparrow$} & R@10 \small{$\uparrow$} & MdR \small{$\downarrow$} & MnR \small{$\downarrow$} & R@1 \small{$\uparrow$} & R@5 \small{$\uparrow$} & R@10 \small{$\uparrow$} & MdR  \small{$\downarrow$} & MnR \small{$\downarrow$} \\
    \hline
    {\color{cGreen}{\color{cGreen}\textit{\small{CLIP-ViT-B/32}}}} & & & & & & & & & & \\ 
    X-Pool~\cite{gorti2022x} & 44.6 & 73.2 & 82.0 & 2.0 & 15.4 &60.0 & 90.0 & 95.0 & 1.0 & 3.8   \\
    DiffusionRet~\cite{jin2023diffusionret}  & 46.7 & 74.7 & 82.7  & 2.0 & 14.3 & -- &-- & --&-- & --  \\
    UATVR~\cite{fang2023uatvr} & 43.1 & 71.8 & 82.3 & 2.0  & 15.1  & 61.3 & 91.0 & 95.6 & 1.0 & 3.3  \\
    CLIP-ViP~\cite{xue2022clip} & 48.6 & 77.1 & 84.4 & 2.0 & -- &-- &-- &-- &-- &--  \\
    \cellcolor{cGrey}VTC (\small{Ours})  &\cellcolor{cGrey}\textbf{52.2} &\cellcolor{cGrey}\textbf{79.3} &\cellcolor{cGrey}\textbf{87.4} &\cellcolor{cGrey}\textbf{1.0} &\cellcolor{cGrey} \textbf{11.1}&\cellcolor{cGrey}\textbf{64.5} &\cellcolor{cGrey} \textbf{94.0} &\cellcolor{cGrey}\textbf{97.3} &\cellcolor{cGrey}\textbf{1.0} &\cellcolor{cGrey}\textbf{2.8}  \\
    \hline 
    {\color{cGreen}\textit{\small{CLIP-ViT-B/16}}} & & & & & & & & & & \\
    X-Pool~\cite{gorti2022x} & 47.3 & 74.8 & 82.8 & 2.0 & 14.2 &62.6 &91.7 &96.0 &1.0 & 3.4 \\
    UATVR~\cite{fang2023uatvr} & 45.8 & 73.7 & 83.3 & 2.0 & 13.5 & 64.5 & 92.6 & 96.8 & 1.0 & 2.8  \\
    CLIP-ViP~\cite{xue2022clip} & 50.5 & 78.4  & 87.1 & 1.0 & -- & -- & --&-- &-- &--  \\
    \cellcolor{cGrey}VTC (\small{Ours})  &\cellcolor{cGrey}\textbf{54.4} &\cellcolor{cGrey}\textbf{81.7} &\cellcolor{cGrey}\textbf{89.3} &\cellcolor{cGrey}\textbf{1.0} &\cellcolor{cGrey}\textbf{9.1}  &\cellcolor{cGrey}\textbf{67.5} &\cellcolor{cGrey}\textbf{95.1} &\cellcolor{cGrey}\textbf{98.2} &\cellcolor{cGrey}\textbf{1.0}  &\cellcolor{cGrey}\textbf{2.4}  \\
    \bottomrule
\end{tabular}
}
}
\end{center}
\caption{Text-to-video comparisons on DiDeMo~\cite{anne2017localizing} and VATEX~\cite{wang2019vatex}.
Bold denotes the best performance. ``--'': result is unavailable. 
} 
\label{tab: benchmark-VATEX-DiDeMo}
\vspace{-8mm}
\end{table*}

\begin{table}[t]
\begin{center}
\scalebox{0.9}{
\setlength{\tabcolsep}{2.4mm}{
\begin{tabular}{c|ccccc} 
    \toprule 
     Method & R@1\small{$\uparrow$} & R@5\small{$\uparrow$} & R@10\small{$\uparrow$} & MdR\small{$\downarrow$} & MnR\small{$\downarrow$} \\
    \hline 
    {\color{cGreen}{\color{cGreen}\textit{CLIP-ViT-B/32}}} & & & & & \\
    ClipBERT~\cite{lei2021less} &6.7  &17.3  &25.2  &32.0  &149.7\\
    CLIP4Clip~\cite{luo2022clip4clip} &9.9  &27.1  &36.8  &21.0  &85.4\\
    X-Pool~\cite{gorti2022x} &11.2  &28.3  &38.8  &20.0  &82.7\\
    \cellcolor{cGrey}VTC (Ours)  &\cellcolor{cGrey}\textbf{16.3} &\cellcolor{cGrey}\textbf{38.4}&\cellcolor{cGrey}\textbf{50.3} &\cellcolor{cGrey}\textbf{12.0} &\cellcolor{cGrey}\textbf{54.0} \\ 
    \hline 
    {\color{cGreen}\textit{CLIP-ViT-B/16}} & & & & & \\
     CLIP4Clip~\cite{luo2022clip4clip} &16.0 &38.2  &48.5  &12.0  &54.1\\
    X-Pool~\cite{gorti2022x} & 20.7 & 42.5 & 53.5 & 9.0 & 47.4 \\
    \cellcolor{cGrey}VTC (Ours)  &\cellcolor{cGrey}\textbf{28.3} &\cellcolor{cGrey}\textbf{53.8} &\cellcolor{cGrey}\textbf{66.4} &\cellcolor{cGrey}\textbf{4.0} &\cellcolor{cGrey}\textbf{28.2} \\ 
    \bottomrule
\end{tabular}
}
}
\end{center}
\caption{Text-to-video comparisons on Charades~\cite{sigurdsson2016hollywood}.  } 
\label{tab: benchmark-Charades}
\vspace{-12mm}
\end{table}

\begin{table*}[t]
\begin{center}
\scalebox{0.93}{
\setlength{\tabcolsep}{3.3mm}{
\begin{tabular}{ccc|ccccc|ccccc} 
    \hline 
     \multirow{2}{*}{$\mathcal{N}$} & \multirow{2}{*}{${\bf h}$} & \multirow{2}{*}{$\mathbf{\hat{t}}$} & \multicolumn{5}{c|}{Charades Retrieval} & \multicolumn{5}{c}{LSMDC Retrieval} \\
     \cline{4-13}
     & & & R@1 \small{$\uparrow$} & R@5 \small{$\uparrow$} & R@10 \small{$\uparrow$} & MdR \small{$\downarrow$} & MnR \small{$\downarrow$} & R@1 \small{$\uparrow$} & R@5 \small{$\uparrow$} & R@10 \small{$\uparrow$} & MdR  \small{$\downarrow$} & MnR \small{$\downarrow$} \\
    \hline
    \ding{55} & \ding{55}&   \ding{55}  & 12.3 & 33.1 & 46.4 & 14.0 & 61.8 & 25.8 & 44.6 & 54.1 & 8.0 & 46.2  \\
    \ding{51} & \ding{55}&   \ding{55}  & 15.0 & 36.0 & 49.0 & 13.0 & 56.5 & 28.5 & 48.5 & 59.0 & 7.0 & 42.0 \\
    \ding{51} & \ding{51}&   \ding{55}  & 15.5 & 37.0 & 50.0 & 12.0 & 55.3 & 28.8 & 48.8 & 59.3 & 6.0 & 41.5 \\
    \ding{55} & \ding{51}&   \ding{51}  & 16.0 & 37.5 & 50.2 & 12.0 & 54.5 & 29.0 & 49.3 & 58.5 & 6.0 & 41.0 \\
    \cellcolor{cGrey}\ding{51}  & \cellcolor{cGrey}\ding{51} & \cellcolor{cGrey}\ding{51}   &\cellcolor{cGrey}\textbf{16.3} &\cellcolor{cGrey}\textbf{38.4}&\cellcolor{cGrey}\textbf{50.3} &\cellcolor{cGrey}\textbf{12.0} &\cellcolor{cGrey}\textbf{54.0} & \cellcolor{cGrey}\textbf{30.1} &\cellcolor{cGrey}\textbf{49.9} &\cellcolor{cGrey}\textbf{60.3} &\cellcolor{cGrey}\textbf{6.0} &\cellcolor{cGrey}\textbf{39.9}\\
    \hline
    \hline 
     \multirow{2}{*}{$\mathcal{N}$} & \multirow{2}{*}{${\bf h}$} & \multirow{2}{*}{$\mathbf{\hat{t}}$} & \multicolumn{5}{c|}{DiDeMo Retrieval} & \multicolumn{5}{c}{VATEX Retrieval} \\
     \cline{4-13}
     & & & R@1 \small{$\uparrow$} & R@5 \small{$\uparrow$} & R@10 \small{$\uparrow$} & MdR \small{$\downarrow$} & MnR \small{$\downarrow$} & R@1 \small{$\uparrow$} & R@5 \small{$\uparrow$} & R@10 \small{$\uparrow$} & MdR  \small{$\downarrow$} & MnR \small{$\downarrow$} \\
    \hline
    \ding{55} & \ding{55}&   \ding{55}  & 48.4 & 77.6 & 85.1 & 2.0 & 13.7 & 62.4 & 91.5 & 96.4 & 1.0 & 3.7  \\
    \ding{51} & \ding{55}&   \ding{55}  & 50.0 & 78.5 & 86.0 & 1.0 & 12.5 & 63.5 & 92.5 & 96.7 & 1.0 & 3.4 \\
    \ding{51} & \ding{51}&   \ding{55}  & 50.5 & 78.8 & 86.3 & 1.0 & 12.2 & 64.0 & 92.8 & 96.9 & 1.0 & 3.3 \\
    \ding{55} & \ding{51}&   \ding{51}  & 51.0 & 79.0 & 86.5 & 1.0 & 12.0 & 64.2 & 93.0 & 97.0 & 1.0 & 3.2 \\
    \cellcolor{cGrey}\ding{51}  & \cellcolor{cGrey}\ding{51} & \cellcolor{cGrey}\ding{51} & \cellcolor{cGrey}\textbf{52.2} &\cellcolor{cGrey}\textbf{79.3} &\cellcolor{cGrey}\textbf{87.4} &\cellcolor{cGrey}\textbf{1.0} &\cellcolor{cGrey}\textbf{11.1}  &\cellcolor{cGrey}\textbf{64.5} &\cellcolor{cGrey}\textbf{94.0} &\cellcolor{cGrey}\textbf{97.3} &\cellcolor{cGrey}\textbf{1.0}  &\cellcolor{cGrey}\textbf{2.8} \\
    \hline
\end{tabular}
}
}
\end{center}
\caption{Ablation study of text cluster ($\mathcal{N}$), semantic signals (${\bf h}$), and cluster embeddings computed by VTC-Att ($\mathbf{\hat{t}}$) on Charades, LSMDC, DiDeMo, and VATEX datasets.
} 
\label{tab: ablation-large}
\vspace{-8mm}
\end{table*}

\begin{table}[t]
\begin{center}
\scalebox{1}{
\begin{tabular}{ccc|ccccc} 
    \toprule 
    $\mathcal{N}$ & ${\bf h}$ & $\mathbf{\hat{t}}$ & R@1\small{$\uparrow$} & R@5\small{$\uparrow$} & R@10\small{$\uparrow$} & MdR\small{$\downarrow$} & MnR\small{$\downarrow$} \\
    \hline 
    \ding{55} & \ding{55}&   \ding{55} & 44.5 & 71.4 & 81.6 & 2.0 & 15.3 \\
    \ding{51} & \ding{55}&   \ding{55} & 49.2 & 73.5 & 83.0 & 2.0 & 13.8 \\
    \ding{51} & \ding{51} & \ding{55} & 50.3 & 75.9 & 84.0 & 2.0 & 11.3 \\
    \ding{55} & \ding{51}&  \ding{51} & 50.1 & 76.2 & 84.1 & 1.0 & 11.1 \\
    \cellcolor{cGrey}\ding{51}  & \cellcolor{cGrey}\ding{51} & \cellcolor{cGrey}\ding{51}   &\cellcolor{cGrey}\textbf{53.1} &\cellcolor{cGrey}\textbf{78.5}
    &\cellcolor{cGrey}\textbf{88.5} &\cellcolor{cGrey}\textbf{1.0} &\cellcolor{cGrey}\textbf{9.6} \\
    \bottomrule 
    \end{tabular}
}
\end{center}
\caption{Ablation study of text cluster ($\mathcal{N}$), semantic signals (${\bf h}$), and cluster embeddings computed by VTC-Att ($\mathbf{\hat{t}}$) on MSRVTT.} 
\label{tab: ablation}
\vspace{-8mm}
\end{table}

Our model is evaluated on five datasets: MSRVTT \cite{xu2016msr}, LSMDC \cite{rohrbach2015dataset}, DiDeMo \cite{anne2017localizing}, Charades \cite{sigurdsson2016hollywood}, and VATEX \cite{wang2019vatex}. (1) The \textbf{MSRVTT} dataset, a comprehensive video benchmark, includes $10,000$ video clips each with $20$ English sentence annotations, facilitating video retrieval; we use the 1K-A testing split \cite{liu2019use}. (2) The \textbf{LSMDC} dataset, a large-scale movie narrative dataset, contains over $118,081$ clips from $202$ movies, each with natural language descriptions, supporting video understanding and multimodal research; we use $1,000$ videos for testing as per \citet{gabeur2020multi}. (3) The \textbf{DiDeMo} dataset offers $10,464$ short video clips with temporal descriptions; we follow the training/testing protocol of \citet{jin2023diffusionret}. (4) The \textbf{Charades} dataset comprises $9,848$ user-generated videos with $66,500$ temporal annotations across $157$ action classes; we adhere to the split protocol in \citet{lin2022eclipse}. (5) The \textbf{VATEX} dataset, a large-scale multilingual video description dataset, includes over $34,911$ English video clips; we follow the train-test split of \citet{chen2020fine}. Performance is evaluated using Recall at rank {1, 5, 10} (R@1, R@5, R@10), Median Rank (MdR), and Mean Rank (MnR) as per \citet{Wang_2024_CVPR}. Our experiments were conducted on a workstation with 256GB of memory and a single NVIDIA H100 GPU.

\subsection{Comparison Results}
As shown in Table \ref{tab: benchmark-MSRVTT9K-LSMDC}, our VTC method outperforms the baseline methods on both MSRVTT and LSMDC datasets. Specifically, for the MSRVTT dataset, VTC achieves a R@1 of $53.1\%$ and $54.8\%$ using CLIP-ViT-B/32 and CLIP-ViT-B/16, respectively, which are the highest among the compared methods. Similarly, on the LSMDC dataset, VTC attains a R@1 of $30.1\%$ and $30.9\%$, demonstrating its superior retrieval capability.
Table \ref{tab: benchmark-VATEX-DiDeMo} presents the results on the DiDeMo and VATEX datasets. Our VTC method consistently achieves the best performance across all metrics. For instance, on the DiDeMo dataset, VTC achieves a R@1 of $52.2\%$ and $54.4\%$ with CLIP-ViT-B/32 and CLIP-ViT-B/16, respectively. On the VATEX dataset, VTC reaches a R@1 of $64.5\%$ and $67.5\%$, outperforming other methods by a significant margin.
The superior performance on these datasets highlights the robustness of our method in handling diverse video content and complex textual queries. The VTC-Att mechanism's ability to assign appropriate weights to text within clusters ensures that the most semantically relevant text-video pairs are prioritized, leading to improved retrieval outcomes.
On the Charades dataset, as shown in Table \ref{tab: benchmark-Charades}, our VTC method achieves a R@1 of $16.3\%$ and $28.3\%$ with CLIP-ViT-B/32 and CLIP-ViT-B/16, respectively. This demonstrates a notable improvement over the baseline methods, particularly in challenging scenarios where video content involves complex activities and interactions.
The effectiveness of our method on the Charades dataset can be attributed to the comprehensive feature fusion strategy employed by VTC, which leverages the Transformer-based attention mechanism to focus on the most relevant frames within a video sequence. This ensures that the model captures the essential details necessary for accurate video search.

\subsection{Ablation Study}

In this section, we present a comprehensive ablation study to evaluate the impact of different components within our proposed VTC framework on the overall model performance. Specifically, we analyze the contributions of the text cluster ($\mathcal{N}$), semantic signals from Sweeper (${\bf h}$), and the cluster embeddings computed by VTC-Att ($\mathbf{\hat{t}}$). The results of our ablation experiments are summarized in Tables \ref{tab: ablation-large} and \ref{tab: ablation}.

\subsubsection{Impact of Text Clustering ($\mathcal{N}$)}
We utilize randomly sampled texts to create text clusters, which simulate the state of text embeddings in the feature space without undergoing clustering model training. This serves as the control group (\ding{55}) for our experiment.
The inclusion of text clustering ($\mathcal{N}$) significantly enhances the retrieval performance across all datasets. By clustering semantically similar texts, our method (\ding{51}) captures richer contextual information, which aids in more accurate video retrieval. For instance, on the Charades dataset, the R@1 metric improves from $12.3\%$ to $15.0\%$ when text clustering is incorporated, demonstrating its effectiveness in enhancing semantic understanding.

\subsubsection{Semantic Signals from Sweeper (${\bf h}$)}.
We use randomly generated ${\bf h}$ as a control group to simulate the absence of semantic signals from Sweeper. Experimental results show that the experimental group outperforms the control group across all datasets. This is because the random ${\bf h}$ lacks meaningful semantic signals, making it ineffective in guiding the model to identify noise in text clusters. Conversely, the trained Sweeper can semantically distinguish noise, offering effective semantic signals to the model.

\subsubsection{Cluster Embeddings via VTC-Att ($\mathbf{\hat{t}}$)}
We employ the mean of all text embeddings within the text cluster as a control group (\ding{55}), and the group using our VTC-Att strategy as the experimental group (\ding{51}).
The VTC-Att mechanism, which computes cluster embeddings ($\mathbf{\hat{t}}$), further refines the retrieval process by mitigating noise introduced by clustering. This mechanism assigns appropriate weights to text clusters based on their relevance to the video content, thereby enhancing retrieval precision. The results indicate that incorporating VTC-Att yields substantial gains in performance. On the VATEX dataset, the R@1 metric improves from $64.0\%$ to $64.5\%$, underscoring the effectiveness of VTC-Att in noise reduction and semantic alignment.

\subsection{Empirical Study of $\mathcal{K}$}
The hyperparameter $\mathcal{K}$ is pivotal in our approach, determining the text count per cluster. To assess its impact on model performance, we experimented with various datasets, setting $\mathcal{K}$ to 3, 5, 10, 20, and 50. Results are detailed in Tables \ref{tab: K} and \ref{tab: K-MSRVTT9K-LSMDC}.

\textbf{Performance with Different $\mathcal{K}$}.
Results show optimal model performance with $\mathcal{K}$ between 3 and 5. For example, on the MSRVTT-1K dataset, $\mathcal{K}=5$ achieves the highest R@1 of 53.1\%, with improvements in other metrics. This trend is consistent across datasets.

\begin{table}[t]
\begin{center}
\scalebox{1}{
\begin{tabular}{c|ccccc}
            \toprule
        \ $\mathcal{K}$ & R@1\small{$\uparrow$} & R@5\small{$\uparrow$} & R@10\small{$\uparrow$} & MdR\small{$\downarrow$} & MnR\small{$\downarrow$} \\
            \hline
            w/o cluster & 44.2 & 72.1 & 81.5 & 2.0 & 13.3 \\
             3 &  52.3 & 77.8 & 87.9 & 1.0 & 9.8 \\
            \cellcolor{cGrey}5 &\cellcolor{cGrey} \textbf{53.1}&\cellcolor{cGrey}\textbf{78.5}&\cellcolor{cGrey}\textbf{88.5} &\cellcolor{cGrey}\textbf{1.0} &\cellcolor{cGrey}\textbf{9.6} \\
            10 & 52.0 & 78.0 & 88.0 & 1.0 & 10.0 \\
            20 & 49.5 & 75.8 & 85.7 & 2.0 & 10.7 \\
            50 & 44.3 & 73.6 & 82.8 & 3.0 & 12.8 \\
            \bottomrule
            \end{tabular}
}
\end{center}
\caption{Optimized discussion of $\mathcal{K}$-nearest neighbors in text clusters on MSRVTT-1K.}
\label{tab: K}
\vspace{-8mm}
\end{table}

\begin{table*}[t]
\begin{center}
\scalebox{0.93}{
\setlength{\tabcolsep}{3.6mm}{
\begin{tabular}{c|ccccc|ccccc} 
    \hline 
      \multirow{2}{*}{$\mathcal{K}$} & \multicolumn{5}{c|}{Charades Retrieval} & \multicolumn{5}{c}{LSMDC Retrieval} \\
     \cline{2-11}
      & R@1 \small{$\uparrow$} & R@5 \small{$\uparrow$} & R@10 \small{$\uparrow$} & MdR \small{$\downarrow$} & MnR \small{$\downarrow$} & R@1 \small{$\uparrow$} & R@5 \small{$\uparrow$} & R@10 \small{$\uparrow$} & MdR  \small{$\downarrow$} & MnR \small{$\downarrow$} \\
    \hline
    w/o cluster & 12.3 & 34.7 & 45.6 & 16.0 & 58.7 & 25.4 & 44.7 & 54.9 & 11.0 & 44.8  \\
    3 & 16.0 & 38.1 & 49.9 & 12.0 & 54.2 & 29.4 & 49.6 & 60.1 & 6.0 & 40.2 \\
    \cellcolor{cGrey}5 & \cellcolor{cGrey}\textbf{16.3} &\cellcolor{cGrey}\textbf{38.4}&\cellcolor{cGrey}\textbf{50.3} &\cellcolor{cGrey}\textbf{12.0} &\cellcolor{cGrey}\textbf{54.0} & \cellcolor{cGrey}\textbf{30.1} &\cellcolor{cGrey}\textbf{49.9} &\cellcolor{cGrey}\textbf{60.3} &\cellcolor{cGrey}\textbf{6.0} &\cellcolor{cGrey}\textbf{39.9}\\
    10 & 15.8 & 38.2 & 49.8 & 12.0 & 54.9 & 29.4 & 49.1 & 59.7 & 6.0 & 40.7 \\
    20 & 14.6 & 37.4 & 48.3 & 13.0 & 55.6 & 29.2 & 47.3 & 57.5 & 8.0 & 41.1 \\
    50 & 13.3 & 36.4 & 47.6 & 14.0 & 56.9 & 27.5 & 46.4 & 56.2 & 10.0 & 43.0 \\
    \hline
    \hline 
     \multirow{2}{*}{$\mathcal{K}$} & \multicolumn{5}{c|}{DiDeMo Retrieval} & \multicolumn{5}{c}{VATEX Retrieval} \\
     \cline{2-11}
      & R@1 \small{$\uparrow$} & R@5 \small{$\uparrow$} & R@10 \small{$\uparrow$} & MdR \small{$\downarrow$} & MnR \small{$\downarrow$} & R@1 \small{$\uparrow$} & R@5 \small{$\uparrow$} & R@10 \small{$\uparrow$} & MdR  \small{$\downarrow$} & MnR \small{$\downarrow$} \\
    \hline
     w/o cluster & 48.1 & 74.3 & 83.3 & 4.0 & 15.0 & 59.9 & 89.5 & 92.3 & 3.0 & 5.4  \\
     3& 51.7 & 79.1 & 86.9 & 1.0 & 11.5 & 64.0 & 93.7 & 97.1 & 1.0 & 3.2 \\
     \cellcolor{cGrey}5 & \cellcolor{cGrey}\textbf{52.2} &\cellcolor{cGrey}\textbf{79.3} &\cellcolor{cGrey}\textbf{87.4} &\cellcolor{cGrey}\textbf{1.0} &\cellcolor{cGrey}\textbf{11.1}  &\cellcolor{cGrey}\textbf{64.5} &\cellcolor{cGrey}\textbf{94.0} &\cellcolor{cGrey}\textbf{97.3} &\cellcolor{cGrey}\textbf{1.0}  &\cellcolor{cGrey}\textbf{2.8} \\
     10& 50.9 & 78.5 & 86.6 & 1.0 & 11.9 & 63.3 & 93.3 & 96.5 & 1.0 & 3.6 \\
     20& 50.4 & 76.3 & 86.0 & 2.0 & 14.1 & 62.1 & 92.2 & 95.7 & 1.0 & 4.3 \\
     50& 49.4 & 75.3 & 84.4 & 4.0 & 14.9 & 61.1 & 91.0 & 93.4 & 2.0 & 5.7 \\
    \hline
\end{tabular}
}
}
\end{center}
\caption{Discussion of $\mathcal{K}$-nearest neighbors in text clusters on the Charades, LSMDC, DiDeMo, and VATEX datasets.
} 
\label{tab: K-MSRVTT9K-LSMDC}
\vspace{-8mm}
\end{table*}

\textbf{Effects of Larger $\mathcal{K}$}.
Increasing $\mathcal{K}$ beyond 10 leads to performance decline due to excessive noise in larger clusters. These clusters include less semantically similar texts, obscuring meaningful relationships and reducing the model's retrieval accuracy.

\begin{table*}[t]
\begin{center}
\setlength{\tabcolsep}{8mm}{
\begin{tabular}{c|cc} 
    \hline 
    Query & Text Cluster & Matching Score \\
    \hline
     \multirow{3}{*}{a cat and a monkey are playing} & a monkey jumping by a cat & 0.6 \\
      & a bird plays with a cat while music plays & 0.1 \\
      & a cat plays with a small child & 0.2 \\
    \hline
     \multirow{3}{*}{A man driving a car} & a man driving a car in the car & 0.8 \\
      & a man is driving & 0.7 \\
      & a car almost driving into another car & 0.3 \\
    \hline
\end{tabular}
}
\end{center}
\caption{Queries from MSRVTT dataset. The Matching score given by the Sweeper to each text in the text cluster.} 
\label{tab: case_study}
\end{table*}

\begin{figure*}
	    \centering
	    \includegraphics[scale=0.53]{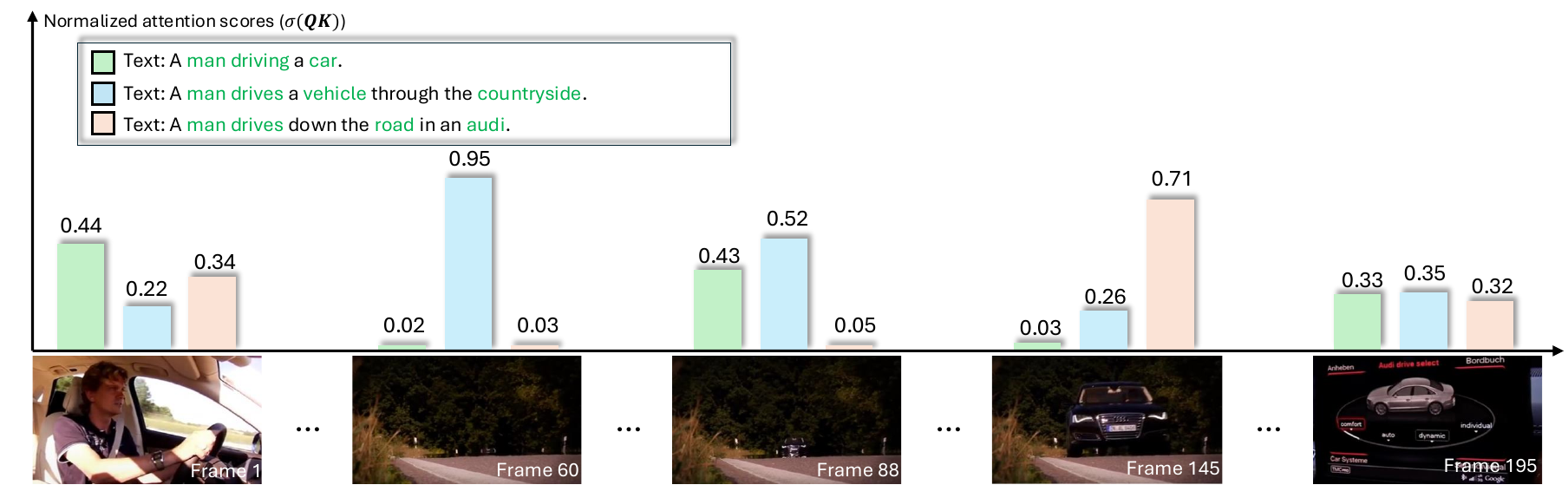}
	    \caption{The visualization of $\sigma({\bf QK})$, with an instance derived from the MSRVTT dataset, depict the attention scores for each frame in the video concerning the texts within the text cluster.}
	    \label{fig:case_study}
\end{figure*} 

\textbf{Guidelines for $\mathcal{K}$ Selection}.
We recommend $\mathcal{K}<10$ to balance semantic capture and noise reduction. This aligns with clustering algorithms' principles \cite{bindra2017detailed}, which group highly similar elements.
Setting $\mathcal{K}$ too large may be detrimental to the model. For instance, if we set $\mathcal{K}$ to the maximum value, which is the size $N$ of the dataset, this would result in text clusters encompassing all samples, which is meaningless for model training.

\subsection{Case Study}
To demonstrate the effectiveness of our Sweeper module in detecting noise within text clusters, we analyzed two queries and their associated text clusters from the dataset. Table \ref{tab: case_study} illustrates that text clusters often contain noise, deviating from the original query's intent. For instance, in the query ``a cat and a monkey are playing'', only ``a monkey jumping by a cat'' closely matches the query, resulting in a higher matching score, while other texts score lower due to semantic differences. The text cluster for the query ``A man driving a car'' is also similar. This confirms that our trained Sweeper efficiently identifies noise in text clusters.

Additionally, to assess the VTC-Att's capability to focus on individual texts using video frame information, we visualized its attention scores, as shown in Figure \ref{fig:case_study}. The figure reveals that frames relevant to the text's semantics receive higher attention scores. For example, the text ``A man drives down the road in an Audi'' has a notably higher attention score for Frame $145$, which depicts an Audi on the road, aligning with the text's meaning. Similar patterns are observed across other frames, validating that VTC-Att effectively identifies text aligned with video semantics, enhancing model focus on video-related content and improving performance.

\section{Conclusion}
This paper introduces a novel framework for video retrieval and proposes an innovative method named VTC (Video-Text Clustering). 
Unlike traditional video retrieval approaches that often struggle with semantic misalignment, our proposed methodology expands query semantics through a sophisticated clustering mechanism, 
effectively bridging the semantic gap between concise textual queries and complex videos composed of multiple image frames.
To address the challenge of additional noise inevitably introduced by the clustering process, we developed the Sweeper module—a specialized component designed to extract and refine semantical signals between the query and clustered text representations. This module significantly enhances the signal-to-noise ratio in the retrieval process, ensuring more accurate matches between queries and video content.
Furthermore, we implemented the VTC-Att mechanism, which synergistically combines the semantical signals from the Sweeper module with video frame signals to extract effective information within the text cluster. This integration ensures that model training remains robust against noise interference, maintaining high performance even with complex query-video pairs.
Comprehensive experimental evaluations across multiple benchmark datasets further validate our approach, showing that the VTC framework consistently outperforms current state-of-the-art baselines on standard metrics. These results underscore the effectiveness of our semantic expansion strategy and noise reduction techniques in advancing the field of video retrieval.

\section*{Acknowledgments}
We are grateful to Huihui Bai and Yunlai Hao for their emotional support during our work, which helped reduce our stress.
We wish all the authors good health and success in all their endeavors.
We sincerely thank the reviewers for their dedicated work and careful evaluation of our manuscript.

\bibliographystyle{ACM-Reference-Format}
\bibliography{sample-base}

\end{document}